\documentclass{article}
\usepackage{graphicx}
\usepackage{epsfig}

\usepackage{pstricks}

\usepackage{amsfonts,amsmath}

\usepackage{amssymb}
\date{\today}

\newcommand{\insertplot}[5]{\begin{figure}
 \hfill\hbox to 0.05in{\vbox to #5in{\vfill
 \inputplot{#1}{#4}{#5}}\hfill}
 \hfill\vspace{-.1in}
 \caption{#2}\label{#3}
 \end{figure}}
 \newcommand{\inputplot}[3]{
 \special{ps: plotfile #1}
}
\newcounter{fig}   
 
\newcommand{\vphi}{\varphi}

\newcommand{\be}{\begin{equation}}
\newcommand{\ee}{\end{equation}}
\newcommand{\bea}{\begin{eqnarray}}
\newcommand{\eea}{\end{eqnarray}}

\usepackage{etoolbox}

\makeatletter
\patchcmd{\maketitle}{\@fnsymbol}{\@alph}{}{}  
\makeatother

\newcommand{\diff}{\mathrm{d}}

\begin{document}

\title{Spinning-Charged-Hairy Black Holes in 5-d Einstein gravity} 
\author{
{\large Y. Brihaye}\thanks{yves.brihaye@umons.ac.be}~$^{\dagger}$, {\large L. Ducobu}\footnote{ludovic.ducobu@umons.ac.be}~$^{\dagger}$
{\large } 
\\ 
\\
$^{\dagger}${\small Physique-Math\'ematique, Universit\'e de
Mons, Mons, Belgium}
\\
}
\maketitle
\begin{abstract}
The spinning-hairy black holes that occur in Einstein gravity supplemented by a doublet
of complex scalar fields are constructed within an extension of the model by a $U(1)$ gauge symmetry
involving a massless vector potential. The hairy black holes then acquire 
an electric charge and a magnetic moment; their domain of  existence  is
discussed in terms of the gauge coupling constant.
\end{abstract}

\section{Introduction} 

One century after the discovery of General Relativity (GR) by Einstein in 1915, the research in 
this topic has never been so intensive and exciting. This is mainly boosted by two antinomic phenomena: 
the direct observation of gravitational waves and the desperately lack of evidence of dark matter.
  For obvious reasons, the main effort is 
realized essentially in four dimensional space-time but there are numerous reasons to study GR in higher dimensions
 (see e.g \cite{Emparan:2008eg} and references therein). 
 Besides looking at our Universe with a String theory or Brane-World point of view, the study of GR
in $d > 4$ dimensions provides an inexhaustible domain of research and a fertile ground for new discoveries and 
innovating techniques.
First the basic Einstein-Hilbert action can be enlarged by a hierarchy of Lovelock actions involving higher 
powers in the Rieman tensor \cite{lovelock}; the first term of which  being the Gauss-Bonnet action.
Second, while the basic classical solutions in 4-d gravity 
 are drastically featureless \cite{Chrusciel:2012jk}
 (the electro vacuum black holes being described by a few macroscopic degrees of freedom),
higher dimensional gravity admit families of new classical solutions, black holes with horizon topologies
other than the sphere \cite{Emparan:2001wn},\cite{Emparan:2008eg}.

Escaping the rigidity of the ``No-hair'' conjecture \cite{Ruffini:1971bza} has been a long fight
which (up to our knowledge) was been broken first time in \cite{luckock_moss} (see also \cite{Bekenstein:1996pn}).
Attracting a lot of interest, families of four-dimensional hairy black holes endowed by scalar hairs
have been constructed in  \cite{Herdeiro:2014goa},\cite{Herdeiro:2014jaa}. 
The model consists of the standard Einstein gravity minimally coupled to a massive, complex, scalar field.
Among  ingredients  entering crucially in this
construction  let us point out that: 
(i) the black hole has to spin sufficiently fast, 
(ii) the solution is synchronized
in the sense that the spinning velocity of the black hole on the horizon coincides with the 
frequency, say $\omega$, parametrizing the harmonic time-dependence of the scalar field,
(iii) the solutions bifurcate for a peculiar sub-family of (hairless) Kerr solutions,
(iv) the solution involves the numerical integration of a set non-linear partial differential equations.

The discovery of these solutions motivated research of similar solutions in different extensions of the model
and/or in different gravity frameworks, see e.g. \cite{Herdeiro:2016gxs}. Also it is natural to emphasize that hairy black holes exist in  higher dimensional  gravity extended by an appropriate matter sector.
In space-time with odd dimensions  the construction of spinning black holes 
is technically simpler since  a suitable ansatz of the metric leads to differential equations (instead of partial derivative ones).
As a consequence, some aspects of hairy black holes  can be studied in such space-time.

In the absence of matter field, the generic spinning black holes of $d>4$ space-time gravity are  the
 Myers-Perry (MP) solutions \cite{Myers:1986un}.
Supplementing the Einstein-Hilbet action by a single complex scalar leads essentially to non-spinning boson stars.
To our knowledge, no parametrization of the metric and matter functions can be implemented to
obtain spinning solutions or black holes.
One possible  key ingredient for obtaining spinning boson stars and black holes
consists in the inclusion of a doublet of complex scalar fields. This has been proposed in \cite{Hartmann:2010pm} for the construction of spinning boson stars in $d = 5$ and, using a similar ansatz for the scalar fields,  spinning black holes in $d = 5$ gravity  were constucted in 
\cite{Brihaye:2014nba} and \cite{Brihaye:2016vkv}. 
One main feature of these results is that, unlike the $d = 4$ case,
the spinning hairy black holes remain decoupled from the family of MP solutions. 
Interestingly, the family of $d=5$ hairy black holes is ``envelopped'' by the boson stars on the one
side and by a family of extremal solution on the second side.
Asymptotically AdS black holes and boson stars
were constructed in \cite{Dias:2011at}  with the same matter contain but with a slightly different  parametrization.
Similar to the $d=4$ case, these black holes bifurcate from the AdS-Myers-Perry solutions.

On the other hand, supplementing gravity by gauge fields often results in families of charged objects
(black holes or solitons) presenting new physically interesting  properties. The oldest and most famous 
example is the occurrence of extremal solutions of Reisner-Nordstrom black holes. Recently,
it was revealed that charged black holes possess the property of superradiance
\cite{Hod:2016kpm,Hod:2017kpt,Hod:2013eea}. 
The coupling of the d=4 hairy black holes solutions of \cite{Herdeiro:2014goa}
to electromagnetism has been investigated in details \cite{Delgado:2016jxq}.

Recently it was shown \cite{Grandi:2017zgz} that, in  d=5, non-spinning hairy black holes exist in the 
Einstein-Gauss-Bonnet-Maxwell theory   provided both -the gauge coupling constant and the Gauss-Bonnet parameters-
are sufficiently large. Several  properties of these solutions have been discussed in \cite{Brihaye:2018nta}.
To our knowledge, the gauge version of d=5 spinning black holes has not yet been studied and this problem
in emphasized in this paper.

The ingredients of the model, the ansatz and the boundary conditions are
presented in the second section. In Sect. 3 we review the properties of the non-hairy solutions: the Myers-Perry
and Reissner-Nordstrom solutions. For completeness, the uncharged spinning hairy black holes are briefly  summarized in Sect. 4 and
the new results are presented in Sect. 5. Because of their rotation, the family of black holes in the full model
 acquire an electric charge as well as a magnetic moment. Finally, in section 6 we summarize our results and conclude.
  

\section{The field equations}
\label{sectmodel}
\subsection{The model}
Following the conventions of \cite{Hartmann:2010pm}, we consider the   action of the self-interacting 
complex doublet scalar field $\Phi$ 
coupled minimally to Einstein gravity in 5 dimensional spacetime and supplemented by a Maxwell field
\begin{equation}
S=\int \left[ \frac{R}{16\pi G}
   -  \left( D_\mu \Phi \right)^\dagger \left( D^\mu \Phi \right) - U( \left|\Phi \right|)
   - \frac{1}{4} F_{\mu \nu} F^{\mu \nu} 
 \right] \sqrt{-g} \diff^5x. \label{action}
\end{equation}
In the first term, $R$ represents the 
curvature scalar and  $G$ the Newton's constant 
(in 5 dimensions).  The second term is the kinetic part of the
scalar field with  $\Phi = (\phi_1,\phi_2)^t$ and  $A^\dagger$ denotes the complex transpose of $A$.  
The third term  $U$ is a self-interaction potential depending on the norm $|\Phi|^2 = \Phi^\dagger \Phi$. 
With this choice, the scalar sector possesses an $U(2)$ global symmetry whose any $U(1)$ subgroup can be gauged.
In this work, we will gauge the diagonal $U(1)\times U(1)$ subgroup. Then the covariant derivative
and Faraday tensor take the form
$$
    D_\mu = (\partial_\mu - i q A_\mu \mathbb{Id}_2 ) \ \ , \ \ F_{\mu \nu} = \partial_\mu A_{\nu} - \partial_\nu A_{\mu},
$$
where $\mathbb{Id}_2$ is the $2\times2$ identity matrix, and $q$ denotes the gauge coupling constant. We will assume $q>0$ since the sign of $q$ can be reabsobed in the gauge fields.

Variation of the action with respect to the metric
leads to the Einstein equations
\begin{equation}
G_{\mu\nu}= R_{\mu\nu}-\frac{1}{2}g_{\mu\nu}R = 8\pi G (T^s_{\mu\nu} + T^v_{\mu\nu})
\  \label{ee} \end{equation}
with stress-energy tensor
\begin{eqnarray}
T^s_{\mu\nu}  =
   (D_\mu \Phi)^\dagger (D_\nu \Phi) 
  + (D_\nu \Phi)^\dagger (D_\mu \Phi)-  \frac{1}{2} 
   g_{\mu\nu} \bigg[ (D_\alpha \Phi)^\dagger (D_\beta \Phi)
  + (D_\beta \Phi)^\dagger (D_\alpha \Phi)\bigg] g^{\alpha\beta}- 
  g_{\mu\nu}  U(|\Phi|) 
  , \label{tmunu}
\end{eqnarray}
\begin{eqnarray}
T^v_{\mu\nu} =  F_{\mu \alpha } F_{\nu}^{\phantom{\nu}\alpha} - \frac{1}{4} g_{\mu\nu} F_{\alpha \beta} F^{\alpha \beta} \ \ ,
\end{eqnarray}
where the upper-script $s$ stand for \emph{scalar} while the upper-script $v$ stand for \emph{vector}, \emph{i.e.} the Maxwell field.

The variation with respect to the matter fields leads respectively to the equations,
\begin{eqnarray}
& &\frac{1}{\sqrt{-g}}
D_\mu\left(\sqrt{-g} D^\mu \Phi \right) = 
 \frac{\partial U}{\partial |\Phi|^2} \Phi \ \ , \ \ \frac{1}{\sqrt{-g}} \partial_{\mu} \sqrt{-g} F^{\mu \nu} = J^{\nu}
 \ \ , \ \ J^{\nu} = i q ((D^{\nu} \Phi)^\dagger \Phi - \Phi^\dagger (D^{\nu} \Phi))
 . \label{feqH} 
 \end{eqnarray}

\subsection{The Ansatz}
While rotating EKG black holes will generically possess two independent
angular momenta and a more general topology
of the event horizon 
we restrict here to configurations with
equal-magnitude angular momenta and a spherical horizon topology.

Thus the solutions possess bi-azimuthal symmetry,
implying the existence of three commuting Killing vectors,
$\xi = \partial_t$, $\eta_1=\partial_{\varphi_1}$, 
and $\eta_2=\partial_{\varphi_2}$.

A suitable metric ansatz in this case reads 
\begin{eqnarray}
\label{metric}
&&\diff s^2 = \frac{\diff r^2}{f(r)}
  +   g(r) \diff \theta^2
+h(r)\sin^2\theta \left( \diff \varphi_1 -W(r)\diff t \right)^2
+h(r)\cos^2\theta \left( \diff \varphi_2 -W(r)\diff t \right)^2 ~~{~~~~~}
\\
\nonumber
&&{~~~~~~}+(  g(r)-h(r))\sin^2\theta \cos^2\theta(\diff \varphi_1 -\diff \varphi_2)^2
-b(r) \diff t^2,
\end{eqnarray}
where $\theta  \in [0,\pi/2]$, $(\varphi_1,\varphi_2) \in [0,2\pi]$,
and $r$ and $t$ denote the radial and time coordinate, respectively.

For such solutions the isometry group is enhanced from $\mathbb{R}_t \times U(1)^{2}$
to $\mathbb{R}_t \times U(2)$, where $\mathbb{R}_t$ denotes the time translation.
  This symmetry enhancement allows to factorize the angular dependence
and thus leads to ordinary differential equations.

Completing the metric (\ref{metric}), the scalar field is taken in the form used in \cite{Hartmann:2010pm}~:
\begin{equation}
 \Phi = F(r) e^{i \omega  t} 
 \left( \begin{array}{c} 
   \sin\theta  e^{i \vphi_1} \\ \cos\theta  e^{i \vphi_2} 
 \end{array} \right) 
 , \label{phi} 
\end{equation} 
where the frequency $\omega$ parametrized the harmonic time-dependence. 
For the scalar field potential we restrict our study to the simplest case 
   \begin{eqnarray}
   U(|\Phi|) =\mu^2 \Phi^\dagger \Phi  =\mu^2 F(r)^2
   \end{eqnarray}
where $\mu$ corresponds to the scalar field mass.

Finally the electromagnetic potential is choosen in the  form 
\begin{equation}
A_{\mu} \diff x^{\mu} = V(r) \diff t + A(r) (\sin^2(\theta) \diff \vphi_1 + \cos^2(\theta) \diff \vphi_2 )
\end{equation} 
which turns out to be consistent with the symmetries of the metric and scalar fields.
The whole ansatz  leads to a consistent
set of differential equations for the radial functions $f,b,h,g,W,V,A$ and $F$.

Without fixing a metric gauge, a straightforward computation
leads to the following reduced action for the system
 \begin{eqnarray}
\label{Leff}
A_{\rm eff}=\int \diff r \diff t ~L_{\rm eff},~~~{\rm with~~~~}
L_{\rm eff}=L_g+16 \pi G  (L_{s}+ L_{v}),
\end{eqnarray}
with
\begin{eqnarray}
L_{g}&=&
\sqrt{\frac{fh}{b}}
\bigg(
b'g'+\frac{g}{2h}b'h'+\frac{b}{2g}g'^2+\frac{b}{h}g'h'+\frac{1}{2}gh W'^2+\frac{2b}{f}(4-\frac{h}{g})
\bigg),
\\
 L_{s}&=& g\sqrt{\frac{bh}{f}}
\left(
fF'^2+(\frac{2}{g}+\frac{(1-q A)^2}{h}-\frac{(\omega-W+q(V+WA))^2}{b}+\mu^2)F^2
\right) ,
\label{leffmat}
\\
L_{v} &=& g\sqrt{\frac{bh}{f}}  \left( \frac{2 A^2}{g^2} + \frac{f}{2 h} (A')^2 - \frac{f}{2b}(V'+W A')^2 \right) 
\end{eqnarray}
 where a prime denotes a
derivative with respect to $r$.

It can be checked  that the full Einstein equations (\ref{ee}) are recovered
by taking the variation
of $A_{\rm eff}$ with respect to $h$, $b$, $f$, $g$ and $W$.
The Klein-Gordon and Maxwell equations are  found by taking the variation
with respect to $F$, $V$ and $A$.

The metric gauge freedom can be fixed afterwards, leading to a system of seven independant
equations plus a constraint which is
a consequence of the other equations.
For the construction of the solutions, we have fixed the metric gauge by taking  
\begin{eqnarray}
g(r)=r^2 
\end{eqnarray}
consistently with the standard analytic form of the Myers-Perry solution.
Appropriate combinations of the equations can be
used in such a way that only the first derivative of $f$ appears in the system. Accordingly,
the equation  for $f(r)$ is a first order equation while the equations of the six other
fields are of the second order.

\subsection{Asymptotics}
The submanifold of space-time characterized by a fixed value of the radial and time coordinates, $r=r_H>0$ and $t=t_0$ respectively,
is a squashed $S^3$ sphere.
Imposing the horizon of the metric by means of the conditions 
$f(r_H)=b(r_H) =  0$ therefore leads to black holes with the same horizon topology.

Restricting to nonextremal solutions, the following expansion
holds near the event horizon:
\begin{eqnarray}
\label{c1}
&&f(r)=f_1(r-r_H)+  O(r-r_H)^2,~~h(r)=h_H+ O(r-r_H),
\\
\nonumber
&&
b(r)=b_1(r-r_H)+O(r-r_H)^2,~~w(r)=\Omega_H+w_1(r-r_H)+O(r-r_H)^2,
\\
\nonumber
&&V(r) = V_H + V_1(r-r_H) + O(r-r_H)^2,~~A(r) = A_H + V_1(r-r_H) + O(r-r_H)^2,
\\
\nonumber
&&F(r)=F_0+F_1(r-r_H)+  O(r-r_H)^2.
\end{eqnarray}

A straightforward calculation gives the following asymptotic expansion
for the solution 
\begin{eqnarray}
\nonumber
&&b(r)=
1
+\frac{{\cal U}}{r^2}+ \cdots,
~~f(r)=
1
+\frac{{\cal U}}{r^2}+\cdots,
~~h(r)=r^2
+\frac{{\cal V}}{r^2}+\cdots ,
W(r)=\frac{{\cal W}}{r^4}+\cdots,
\\
\label{inf1}
&&
V(r) = V_0 + \frac{q_e}{r^2}+ \cdots , A(r) =  \frac{q_m}{r^2}+ \cdots,
~~
F(r)= c_0\frac{e^{-\sqrt{\mu^2-(\omega-qV_0)^2}}r}{r^{3/2}}+\cdots,~~ 
\end{eqnarray}
which guarantees  Minkowski spacetime background to be
approached at infinity.
In this expansion, ${\cal U},{\cal V},{\cal W}$,$q_e$,$q_m$,$V_0$ and $c_0$ are free parameters
that can be reconstructed from the  result of the numerical integration on $[r_H,\infty[$.

Note that, in the last equation (\ref{inf1}), one can see that the scalar field acquires a squared effective mass $\displaystyle M_{\text{eff}}^2 = \mu^2 - (\omega - q V_0)^2$. The condition
\begin{equation}
	\label{condloc}
	\mu - |\omega-qV_0| > 0
\end{equation}
 should therefore  be obeyed to guarantee bound-state (or localized) solutions.

This condition generalize the bound-state condition known for uncharged solutions : that is $\omega < \mu$, see fig. \ref{massomega} (where we put $\mu=1$ without loss of generality).

\subsection{Boundary conditions}

We will not write explicitely the field equations which are lengthy and non illuminating.
However, one can easily see from (\ref{leffmat}) that the regularity of the solutions at the horizon implies
a generalized synchronization condition between the frequency $\omega$ of the scalar field and
the rotation velocity $\Omega_H \equiv W(r_H)$ of the black hole at the horizon
\begin{equation*}
\omega -  \Omega_H  + q (V_H + A_H \Omega_H) = 0.
\end{equation*}
This condition can be used to determines the frequency $\omega$ in terms of the other parameters.

Because the electric potential $V(r)$ can be shifted up to a redefinition of quantity 
$A_H \Omega_H$ and the frequency $\omega$, we will take advantage of this freedom to set $V_H=0$. The condition would then become
\begin{eqnarray}
\label{synch}
\omega -  \Omega_H  + q A_H \Omega_H = 0.
\end{eqnarray}
This condition generalize the so called ``synchronisation condition'' ($\omega = \Omega_H$) known for the uncharged case. We clearly see here that, for a given value of $\Omega_H \neq 0$, the synchronisation condition hold \textbf{iff} $q = 0$ or $A_H = 0$.
 
The regularity of the equations for the fields $h(r), W(r)$ and the Maxwell equations imply three independent non trivial
conditions to be obeyed at the horizon, we will note them symbolically $\Gamma_h, \Gamma_W, \Gamma_V$;
these are lengthy polynomials in the various fields and their first derivative, we do not write them 
explicitly because they are not illuminating. Summarizing, we have eight conditions at the horizon
\begin{equation}
\label{condinf}
f(r_H)=0 \ , \ b(r_H)=0 \ , \ W(r_H) = \Omega_H \ , \ V(r_H) = 0 \ , \ A(r_H) = A_H \ , \ \Gamma_{h,W,V} = 0 \ \ .
\end{equation} 
where $A_H$ is an undetermined constant.

The boundary value problem is then fully specified by imposing the conditions (\ref{condinf}) at the horizon  and the asymptotic conditions (\ref{inf1}) for $b,h,W,A$ and $F$.

\subsection{Rescaling}
The theory is determined by three parameters~: $G,\mu,q$. The constants $G$ and $\mu$  can be rescaled 
into the matter fields and the radial variable.
We will use rescaled quantities such that $\mu=1$ and $8 \pi G=1$.
The gauge coupling $q$ is therefore the only intrinsic parameter of the model.

The hairy black holes will  be specified by the event horizon radius $r_H$, the horizon velocity $\Omega_H$
and the value $A_H$. 
%

The (constant) horizon angular velocity $\Omega_H$
is defined in terms of the Killing vector
$\chi=\partial/\partial_t+
 \Omega_1 \partial/\partial \varphi_1 + \Omega_2 \partial/\partial \varphi_2 $
which is null at the horizon.
For the solutions within the ansatz (\ref{metric}), the
horizon angular velocities are equal, $\Omega_1=\Omega_2=\Omega_H$.

\subsection{Quantities of interest}
The mass and angular momentum of the solutions are given by
\begin{eqnarray}
\label{global-charges}
M=-\frac{3S_3 }{16\pi G}{\cal U},~~
J=\frac{ S_3 }{8\pi G}{\cal W},
\end{eqnarray}
where $S_3=2\pi^2$ denotes the area of the unit three-dimensional sphere.

The solution is further characterized by the electric charge $Q_e$ and magnetic moment $Q_m$,
these quantities are related to the parameters $q_e$ and $q_m$ by
\begin{equation}
Q_e = \frac{2 S_3}{4 \pi G} q_e \ \ \ , \ \ \ Q_m = \frac{2 S_3 \pi^2}{4 \pi G} q_m \ \ \ .
\end{equation} 
Finally the gyromagnetic ratio  $\tilde{g}$ can be computed~: $\tilde{g} = 2 M Q_m/ (Q_e J)$. 
Other quantities of interest are
the Hawking temperature  $T_H $
and  the  area $\mathcal{A}_H$ of the black hole horizon
\begin{eqnarray}
\label{Temp-rot}
  T_H=\frac{\sqrt{b_1f_1}}{4\pi},~~\mathcal{A}_H= \sqrt{h_H}  r_H^2 S_3.
\end{eqnarray}
The horizon mass and angular momentum $M_H, J_H$ can further be defined 
(see e.g. \cite{Brihaye:2018woc}); these quantities obey a Smarr relation of the form
\begin{equation}
M_H = \frac{3}{2} (\frac{\mathcal{A}_H T_H}{4} + 2 \Omega_H J_H).
\end{equation} 
This has been used as a numerical test of our numerical results.

In order to figure out the way the distribution of the energy of the scalar
field is affected by the electromagnetic field, we will present the energy
density of some solutions. For completeness, we write these formulas 
\begin{equation*}
{\cal E}_v = g \sqrt{\frac{b h}{f}} \bigl(\frac{f}{2bh}(A')^2(b-hW^2) + \frac{2}{g^2}A^2 + \frac{f}{2b} (V')^2 \bigr),
\end{equation*}
\begin{equation*}
{\cal E}_s = g \sqrt{\frac{b h}{f}} \left( f (F')^2 + \mu^2 F^2 + F^2 \left(\frac{2}{g} + \frac{(1-qA)^2}{h}\right) + \frac{F^2}{b} \left[(qV-\omega)^2- W^2(qA-1)^2\right] \right).
\end{equation*}


\section{Non-hairy solutions}
\label{sectnonhairy}
In the absence of the scalar field ($\Phi=0$), two types of solutions exist
that are know in analytic form.
\subsection{Uncharged, spinning black holes}
In the absence of the Maxwell field
one recovers the MP (vacuum) black holes \cite{Myers:1986un}
with equal-magnitude angular momenta.
Expressed in terms of the event horizon radius
and the horizon angular velocity\footnote{This metric is usually expressed
in terms of the mass $M$ and the angular momentum parameter $a$,
with $M=r_H^2/2(1-r_H^2\Omega_H^2)$ and $a=r_H^2\Omega_H$.}
(which are the control parameters in our numerical approach),
this solution reads
\begin{eqnarray}
\nonumber
 &&f(r)=1-\frac{1}{1-r_H^2\Omega^2_H}\left(\frac{r_H}{r}\right)^2
 +\frac{r_H^2\Omega_H^2}{1-r_H^2\Omega_H^2}\left(\frac{r_H}{r}\right)^4,
~~
 h(r)=r^2
 \left(
 1+\left(\frac{r_H}{r}\right)^4\frac{r_H^2\Omega_H^2}{1-r_H^2\Omega_H^2}
  \right),
\\
\label{MP}
&&b(r)=1
 -\left(\frac{r_H}{r}\right)^2
 \frac{1}{1-\left(1-(\frac{r_H}{r})^4\right)r_H^2\Omega_H^2},
 ~~W(r)=(\frac{r_H}{r})^4\frac{ \Omega_H}{1-\left(1-(\frac{r_H}{r})^4\right)r_H^2\Omega_H^2}.
\end{eqnarray}
Therefore, for a MP black hole, the relevant parameters
in the event horizon expansion (\ref{c1}) are
\begin{eqnarray}
\label{g1}
f_1=\frac{2 (1-2r_H^2\Omega_H^2)}{r_H(1-r_H^2\Omega_H^2)},
 ~~
b_1= \frac{2}{r_H}(1-2r_H^2\Omega_H^2),
 ~~
w_1= \frac{4\Omega_H}{r_H}( r_H^2\Omega_H^2-1),
\end{eqnarray}
while the constants ${\cal U},~{\cal V}$ and ${\cal W}$
in the far field expansion (\ref{inf1}) have the following
expression 
\begin{eqnarray}
\label{g2}
 {\cal V}=\frac{r_H^6 \Omega_H^2}{1-r_H^2\Omega_H^2},
 ~~
  {\cal U}=-\frac{r_H^2  }{1-r_H^2\Omega_H^2},
 ~~
  {\cal W}= \frac{r_H^4 \Omega_H }{1-r_H^2\Omega_H^2}.
\end{eqnarray}
The $d=5$ MP black holes with equal-magnitude angular momenta
emerge smoothly from the static Schwarzschild-Tangherlini black hole
when the event horizon velocity $\Omega_H$ is increased from zero.
For a given event horizon radius,
the solutions exist up to a maximal value of
the horizon angular velocity $\Omega_H^{(c)}=1/\sqrt{2}r_H$ for which the black hole becomes extremal ($T_H \to 0$).
Expressed in terms of the mass-energy $M$ and the equal-magnitude
angular momenta $|J_1|=|J_2|=J$,
this bound reads $27 \pi J^2/8G<M^2$.
The extremal solution saturating this bound
has a regular but degenerate horizon. 
\subsection{Charged, non spinning black holes}
In the absence of rotation, charged black holes exist generalizing the Reisner-Nordstrom  solutions 
The equations are considerably simpler since $f(r)=b(r), h(r)=r^2$ and $W(r)=A(r)=0$
while the non trivial fields read 
\begin{equation}
f(r) = 1 - \frac{2M}{r^2} + \frac{Q^2}{3 r^4} \ \ \ , \ \ \ V(r) = V_0 + \frac{Q}{r^2} \ \ , \ \ 
M = \frac{r_H^2}{2} + \frac{Q^2}{6r_H^2}
\end{equation}
These black holes exist for $0 \leq  Q^2 < 3 r_H^4 $ and  becomes extremal in the limit $Q^2 \to 3 r_H^4$.
\section{Uncharged Hairy Black Holes}
\label{sectunchargedhairy}
\begin{figure}[h]
\begin{center}
{\label{mass_omega_1}\includegraphics[width=8cm]{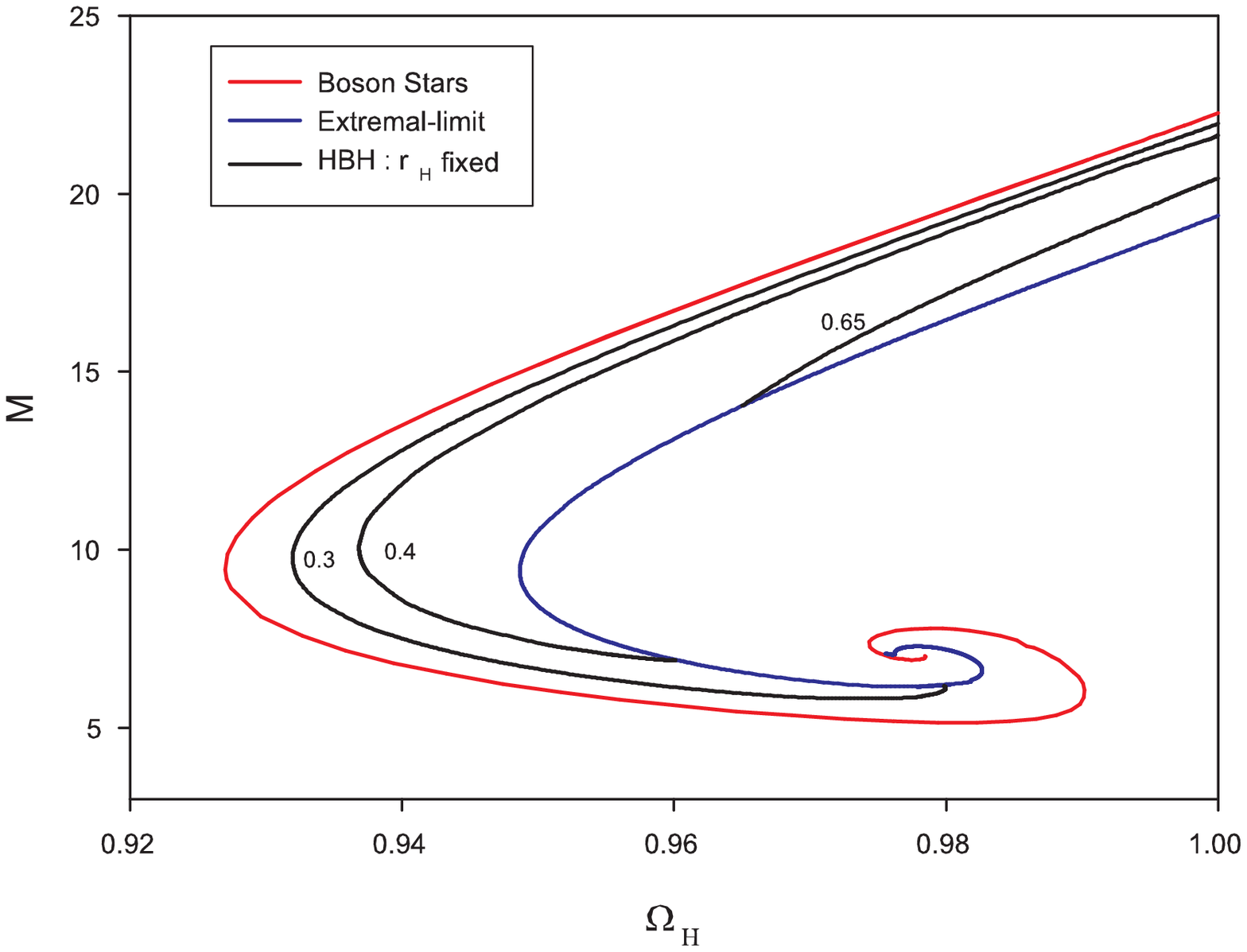}}
{\label{omega_2}\includegraphics[width=8cm]{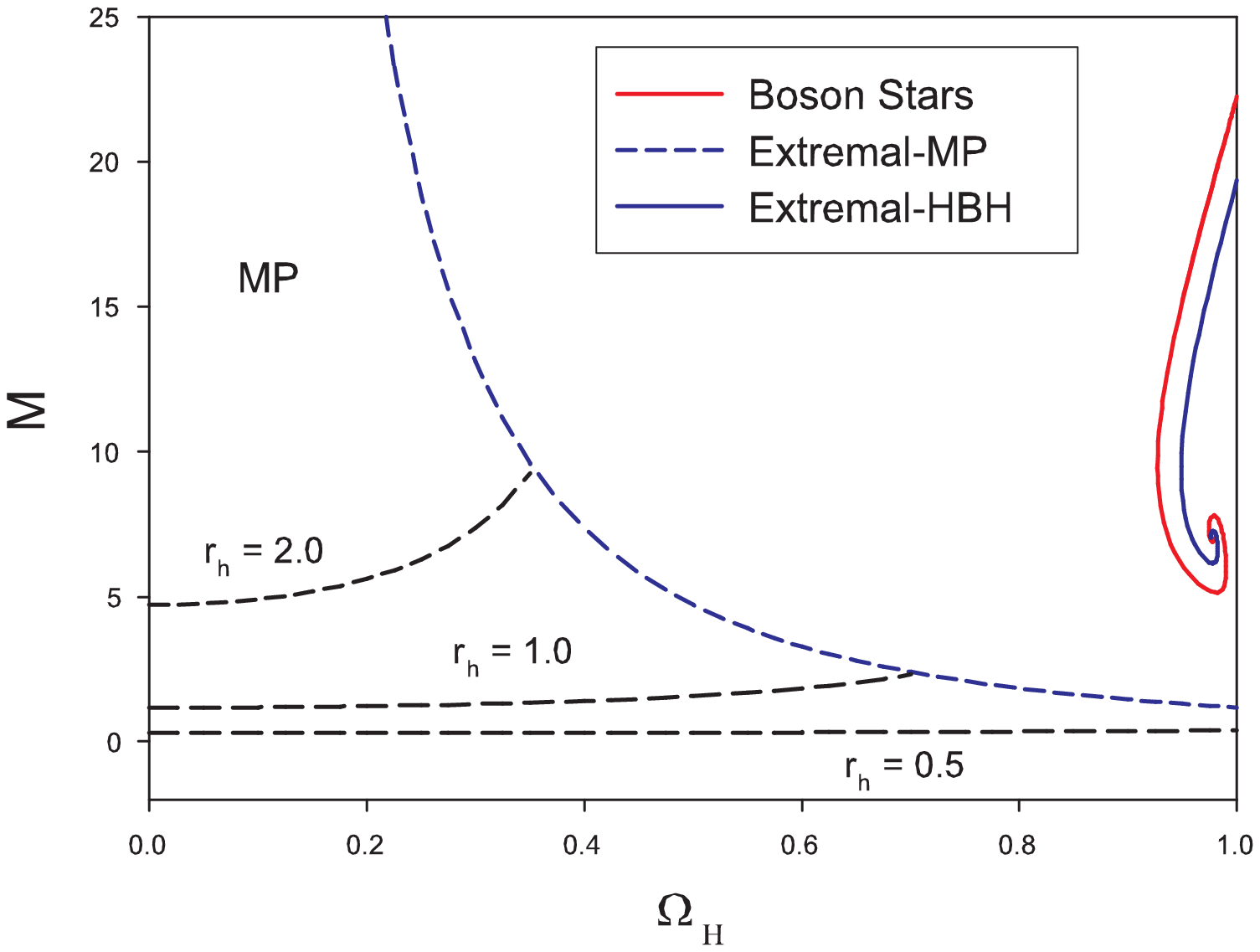}}
\end{center}
\caption{Left: Mass-frequency relation for boson star solutions (red line), for
hairy black holes with three values of $r_H= 0.3,0.4,0.65$ (black lines) and for extremal solutions (blue line). 
Right: Sketch of the domains of Myers-Perry and hairy black holes in the $\Omega_H-M$ plane.
\label{massomega}
}
\end{figure}

In the absence of the electromagnetic field, a family of hairy black holes exist
on a specific domain of the $r_H,\Omega_H$ parameter space. These solutions were obtained in 
\cite{Brihaye:2014nba} but we shortly discuss them again for completenes.
The $\Omega_H-M$ domain of these solutions is represented on Fig. \ref{massomega} (left side) where 
a few values of $r_H$ are supplemented. The solutions exist for $0.924 < \Omega_H < 1$ in the region
 limited by bosons stars (red line, reached in the
limit $r_H \to 0$) and by a family of singular extremal solutions (blue line: $T_H=0$). The
intermediate constitutes the regular black holes. On the right side of Fig. \ref{massomega}
The domain of Myers-Perry and hairy black holes are superposed, demonstating that both families
of solutions are disjoined.
\section{Charged-hairy-solutions}
\label{sectchargedhairy}

Any hairy black holes, e.g. any point of Fig. \ref{massomega} (left side) caracterized by a couple $(r_H, \Omega_H)$
gets  deformed in the presence of an electromagnetic field. 
The purpose of this section is to study the pattern of solutions in dependence of the gauge coupling $q$. 
Unlike the Myers-Perry solutions, the equations with the matter fields have no closed form solutions
and have to be solved numerically.
We used the numerical routine COLSYS \cite{COLSYS} to perform the integration. The radial interval was discretized 
by means of a mesh of $300$ points and the equations were solved with a relative error of less that $10^{-6}$. 

The frequency of the scalar field $\omega$ is fixed via the condition (\ref{synch}). Accordingly, synchronized solutions (that is with $\omega = \Omega_H$) occur \textbf{iff} $q = 0$ or $A_H = 0$. These cases will be discussed separately.

\subsection{Synchronized solutions: Case $q=0$}
This case corresponds to the ungauged model, whose $d=4$ counterpart is studied in details in Sect. 2 of \cite{Delgado:2016jxq}.
In the absence of a direct coupling of the scalar fields to the electromagnetic field (i.e. for $q=0$),
the spinning solution gets deformed through a non-trivial magnetic potential $A(r)$ through the influence of
the rotating space-time. The parameter controling the magnetic potential is $A_H \equiv A(r_H)$. The electromagnetic field vanishes identically for $A_H=0$ and becomes non trivial for $A_H \neq 0$.
\begin{figure}[ht!]
\begin{center}
{\includegraphics[width=8cm]{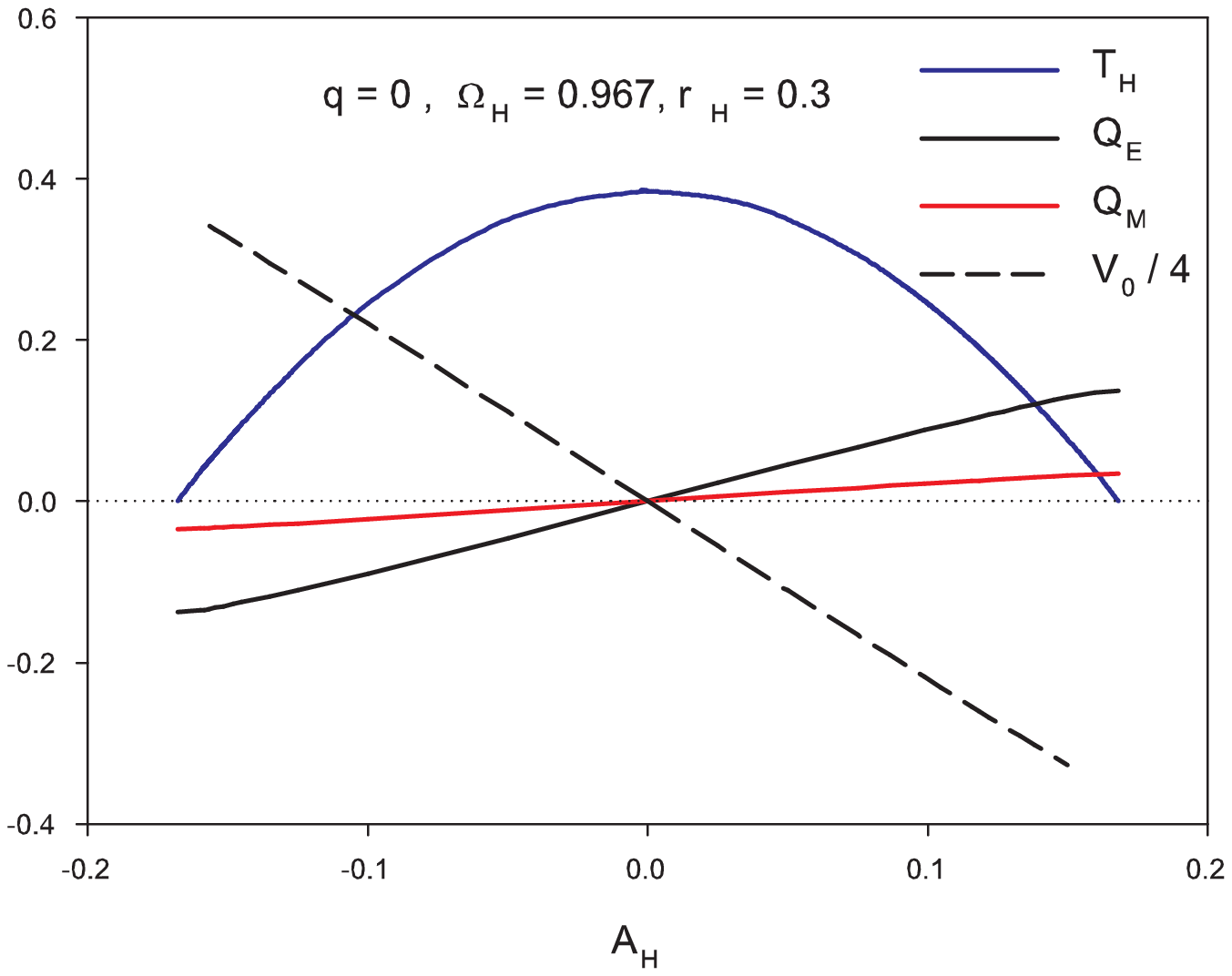}}
{\includegraphics[width=8cm]{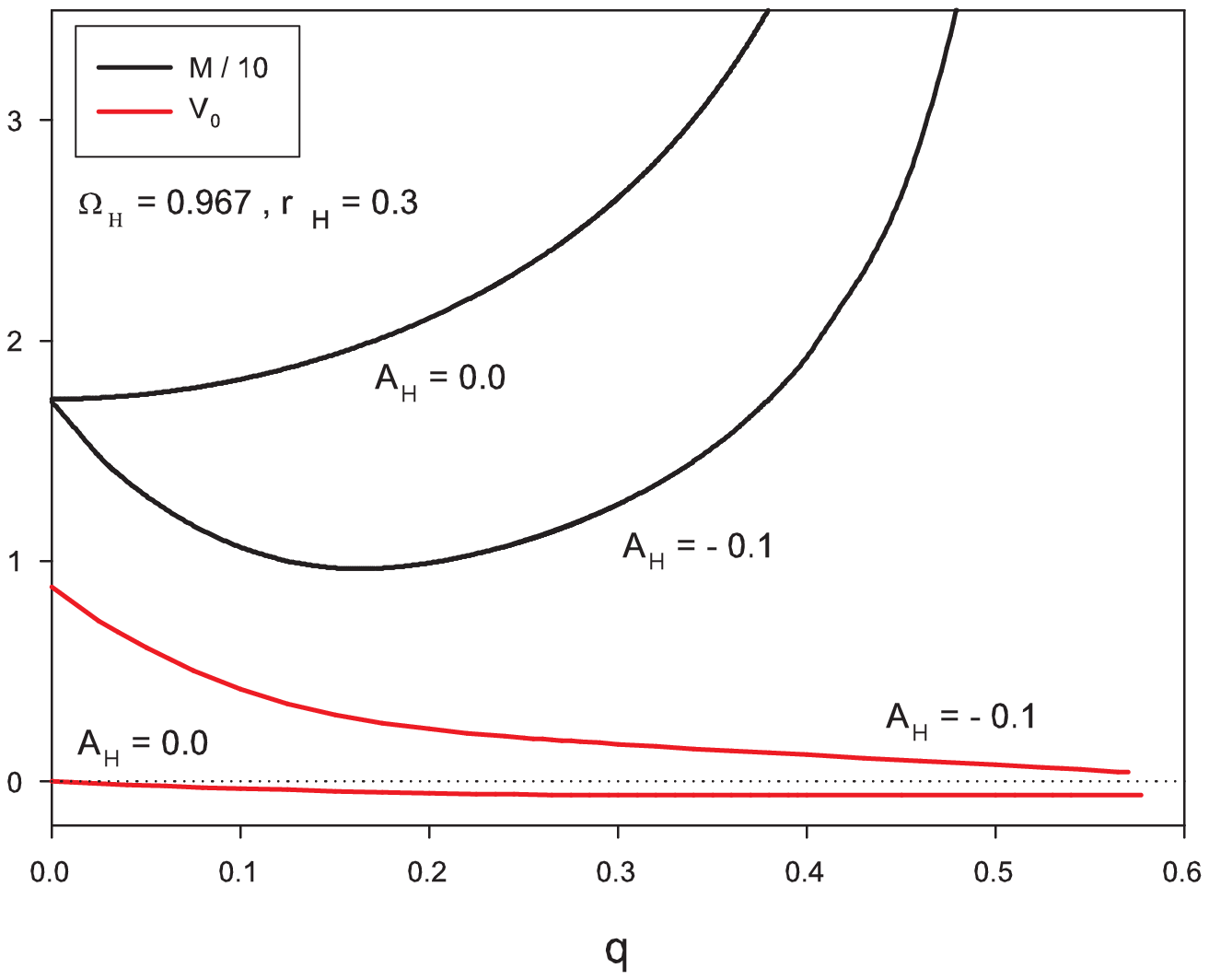}}
\caption{ {\it Left}: Several quantities characterizing the black holes for $q=0$ as functions of the parameter $A_H$.
{\it Right:} The mass $M$ and electric potential $V_0$ as function of $q$ for the solutions corresponding to
$A_H=0.0$ and $A_H=-0.1$.
\label{data_e}
}
\end{center}
\end{figure} 
The effect is  demonstrated on Fig. \ref{data_e} (left side)~: the electric and magnetic parameters $Q_e$, $Q_m$ and the 
electric potential $V_0 \equiv V(\infty)$ are reported as functions of $A_H$. 
The temperature $T_H$ of the black hole is 
supplemented and reveals that a limiting configuration with zero temperature is approached for a maximal value of $|A_H|$. 
In the ungauged case, since $q=0$, the equations simplify drastically. In particular
they become symmetric under $A_{\mu} \rightarrow -A_{\mu}$,
explaining the symmetry of Fig. \ref{data_e} (left) under the sign reversal of $A_H$.
The mass and angular momentum  of these solutions depend  weakly on $A_H$ and have not been reported.
The qualitative features of Fig. \ref{data_e} seem to be generic, it was  checked  
 for a several values of the parameters $r_H$ and $\Omega_H$. 
\subsection{Synchronized solutions~: case $A_H=0$}
\begin{figure}[ht!]
\begin{center}
{\includegraphics[width=8cm]{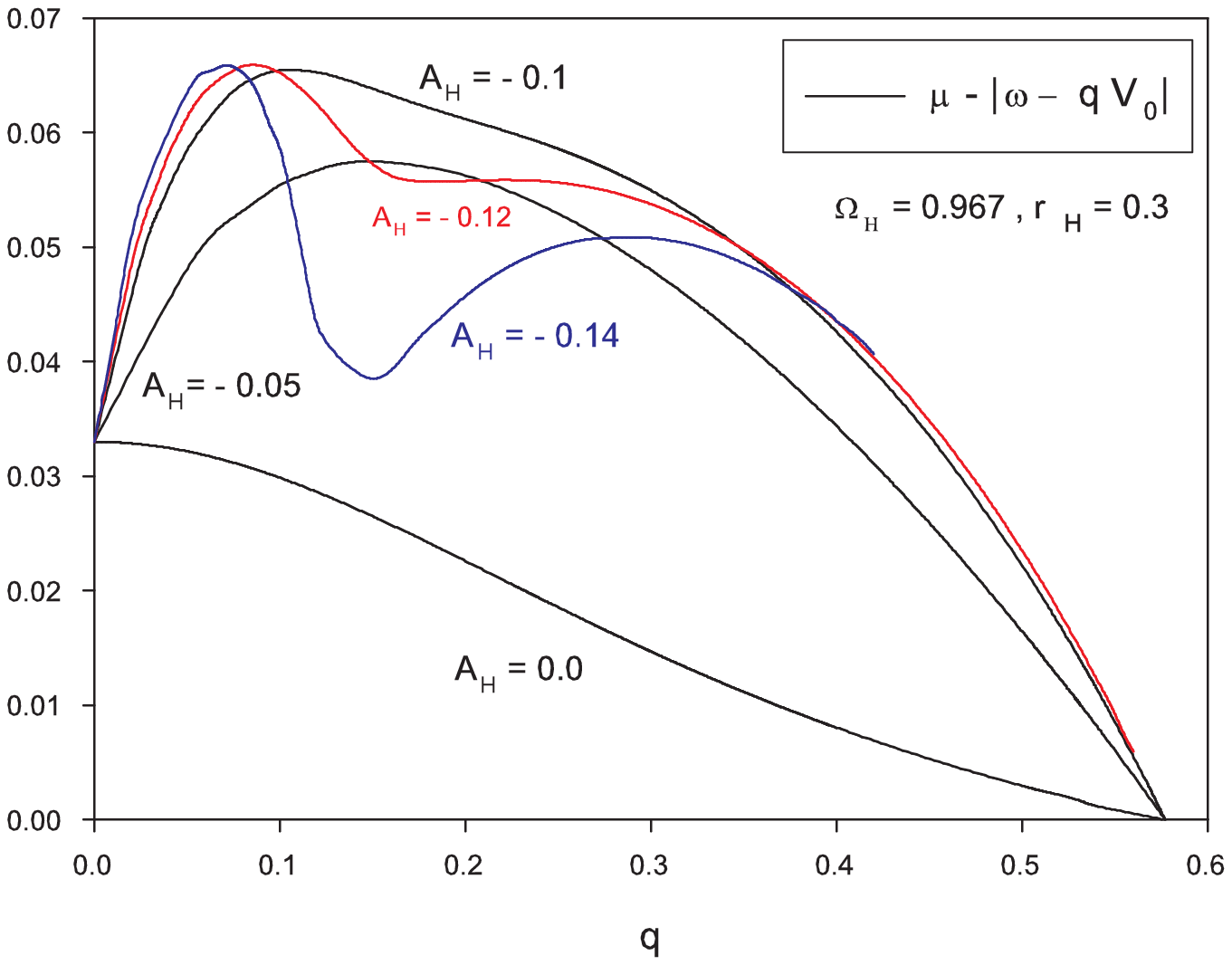}}
{\includegraphics[width=8cm]{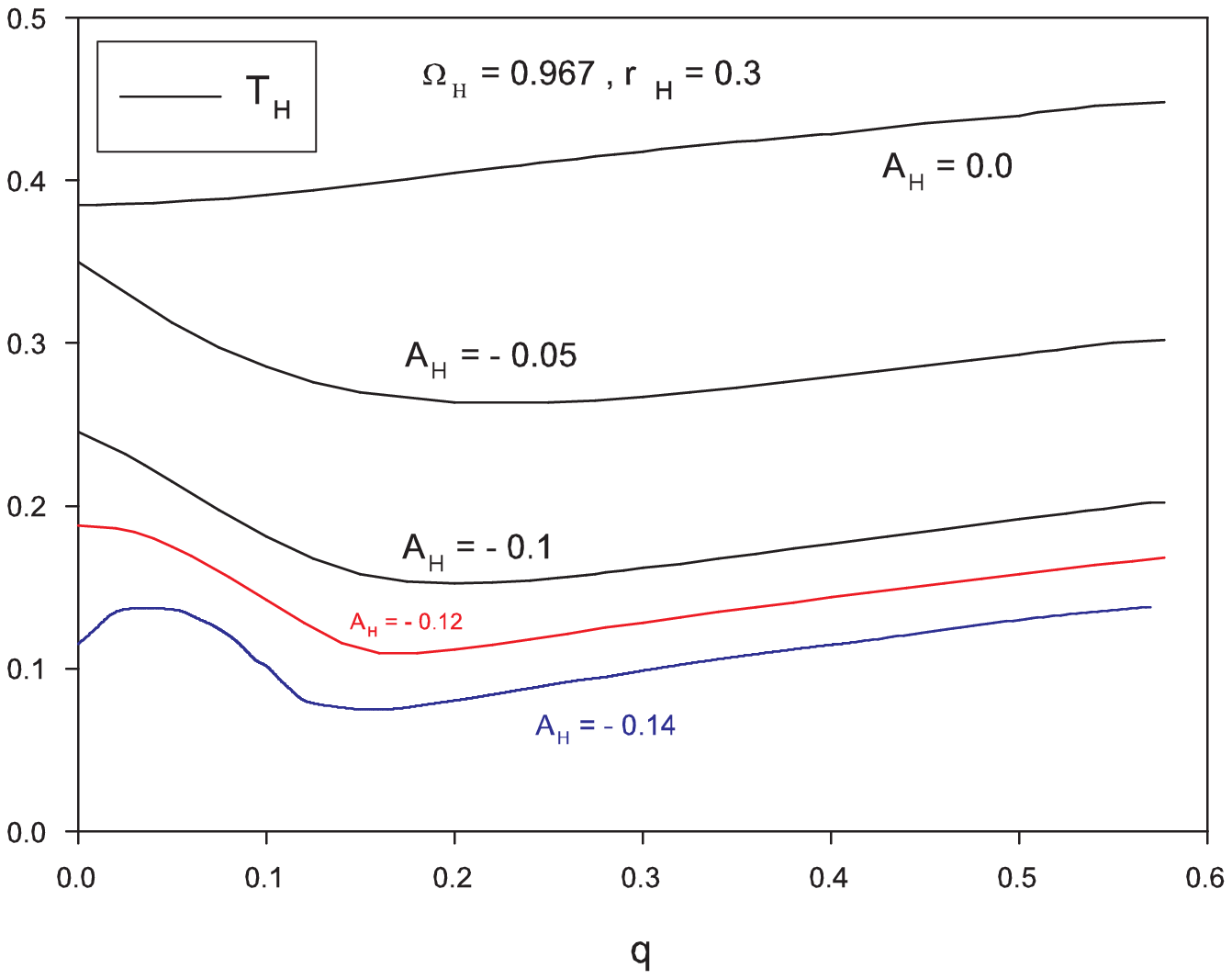}}
\caption{ {\it Left}: The combination $\mu-\left\vert\omega-q V_0\right\vert$ as function of $q$ for several values of $A_H \le 0$. 
{\it Right:} The Temperature for the same set of solutions.
\label{data_a}
}
\end{center}
\end{figure}

Synchronized black holes strongly coupled to the electromagnetic field  are obtained by
increasing gradualy the gauge coupling parameter $q$ while imposing $A_H=0$ as boundary condition.
The electric potential $V_0$ is negative and decreases monotonically while $q$ increases;
this is shown on the right side of Fig.\ref{data_e}.
Our numerical results strongly suggest that the effective squared mass $\mu^2 - (\omega-q V_0)^2$ approaches zero for $q \to q_{\max}$ (see Fig. \ref{data_a} (left side)); as a consequence
localized solution do not exist for $q > q_{\max}$. The value $q_{\max}$ in fact corresponds to the value of 
the coupling constant for which the bound state condition (\ref{condloc}) ceases to be satisfied.
On the example of  Fig. \ref{data_a} , i.e.  with $r_H= 0.3, \Omega_H = 0.967$, we find    $q_{\max} \approx 0.577$.
The temperature corresponding to this set of solutions has been reported on the right side of Fig. \ref{data_a}.

\begin{figure}[ht!]
\begin{center}
{\includegraphics[width=8cm]{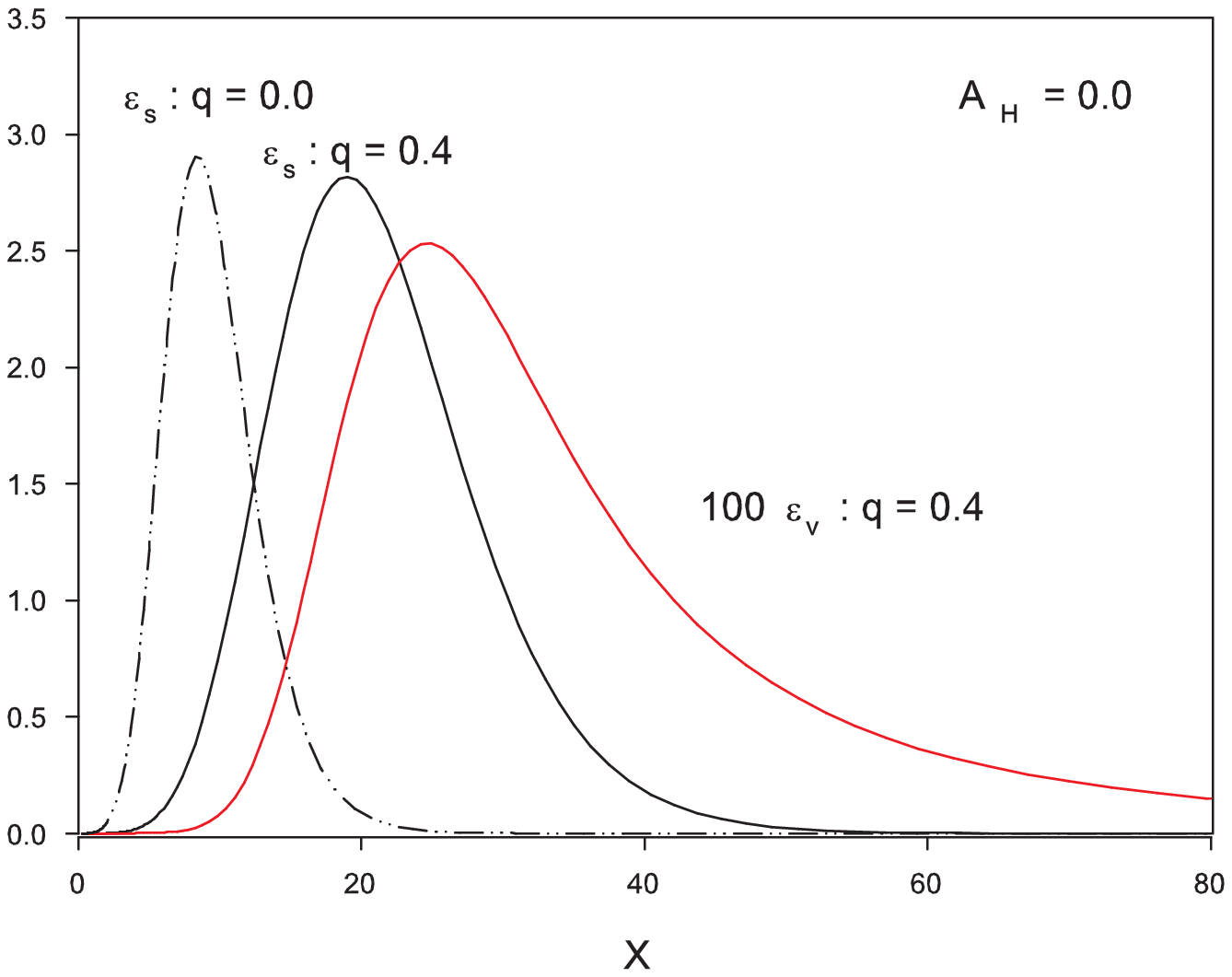}}
{\includegraphics[width=8cm]{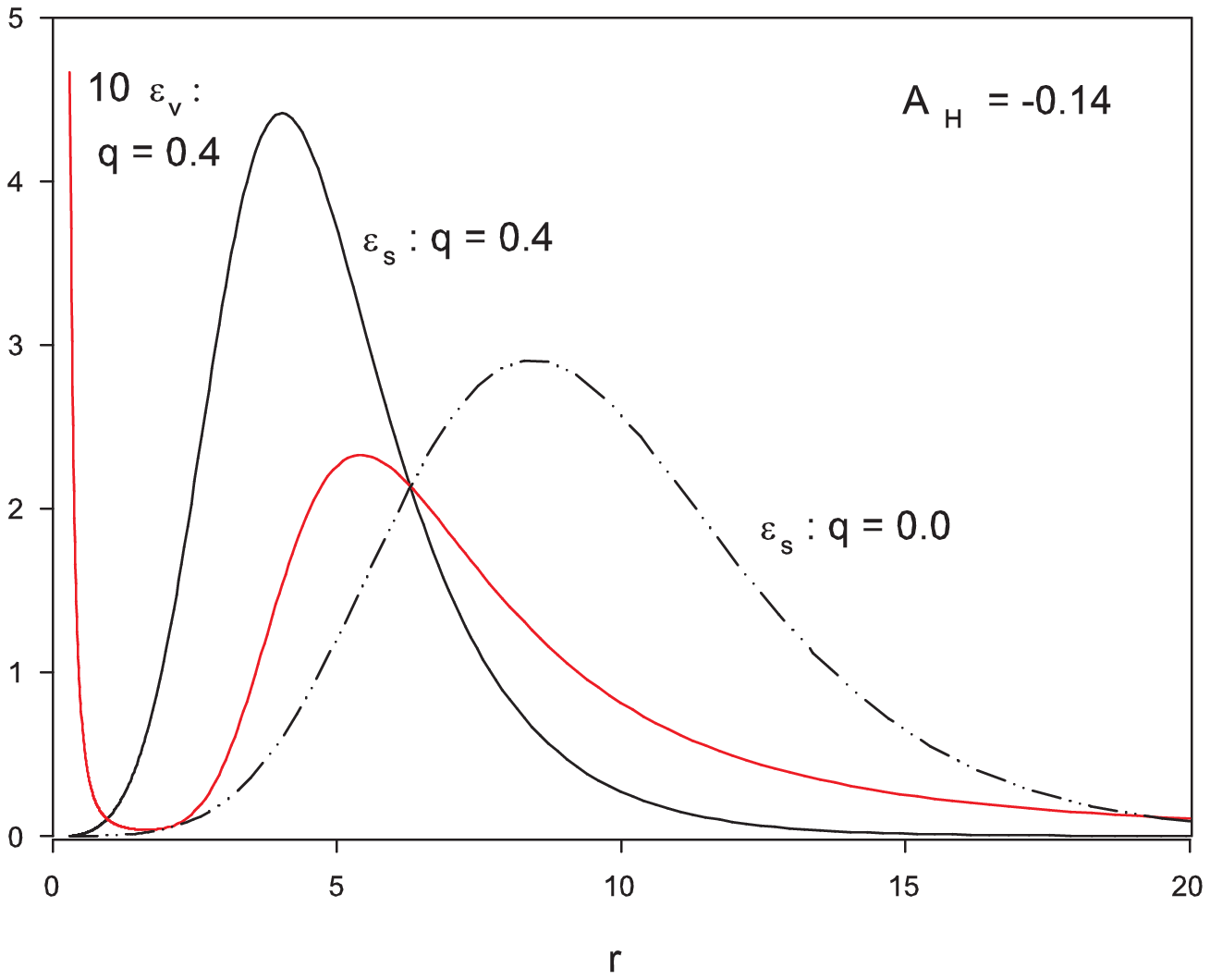}}
\caption{ {\it Left}: The scalar (black) and vector (red) energy density for $q=0$ (dot-dashed) and for $q=0.4$ (solid) for $A_H=0$.
{\it Right:} Idem for $A_H=-0.14$. On both side the dot-dashed curve is the same allowing the reader to compare the plots. Both plots are realized for $r_H=0.3$ and $\Omega_H=0.967$.
\label{energy_dens}
}
\end{center}
\end{figure}

On Fig. \ref{energy_dens} (left side)
the effective energy density of the scalar part and vector part of the matter fields are reported
 for $q=0.0$ and $q=0.4$. 
For $q=0$, the energy density of the vector field is identically null. One can also see that the maximum of the scalar energy density is closer to the horizon in this case compared to the case $q \neq 0$.
This might be interpreted as follows : The case $q=0$ correspond to a uncharged massive scalar field. In this case, the existence of bound state is insured by the gravitational interaction and the scalar ``cloud'' is located at a given distance of the black hole horizon. In the case $q \neq 0$ the scalar field is electrically charged, more precisely every quantum in the ``cloud'' acquire the same charge $q$, and then every quantum would electrically try to grow back each other so that the macroscopic scalar field (the position of the maximum) would be localized further away compared to the uncharged case. This qualitative argumentation can also explain the existence of $q_{\max}$. This correspond to the value of the scalar electric charge for which gravity cease to be strong enough to compensate for the electrostatic repulsion and guarantee a bound state.
\subsection{General case : $A_H \neq 0$, $q \neq 0$}
In the general case, any uncharged hairy solution (with fixed $r_H, \Omega_H$)  lead
to a two-parameter family of charged solutions characterized by the value $A_H$ and the gauge coupling constant $q$. 
This domain of existence of the solutions is largely limited by the bound state condition (\ref{condloc}). 

Setting for definiteness $A_H < 0$, the numerical results indicate that the solutions
stop to be localized for $q > q_{\max}$, i.e.  when the effective frequency becomes imaginary.
These results also strongly suggest that the value $q_{\max}$ is independent of the value $A_H$.
To illustrate this statement, the dependence on $q$ of the quantity appearing in (\ref{condloc}) is reported on Fig. \ref{data_a}
(left side) for several values of $A_H \le 0$. The plot clearly suggest that the value $q_{\max}$ is independent
of $A_H$, although the curves are substantially different for the intermediate values of $q$.
The temperature of these families of solutions are reported on the  right side of  Fig. \ref{data_a}.

For $A_H > 0$, the domain is considerably smaller because, due to the negative sign of $V_0$,
the effective frequency quickly becomes imaginary. We did not study this case in details.

On Fig. \ref{energy_dens} (right side)
we report the effective energy density for the scalar part and vector part
for $A=-0.14$ and for $q=0.0$ and $q=0.4$. Here, the maximum of the scalar energy density is closer to the horizon for $q \neq 0$. This behaviour remains true until $q$ becomes close to $q_{\max}$. While approaching $q_{\max}$ the maximum of the scalar energy density would become more and more distant from the horizon until the bound state condition stop to be satisfied.

The interpretation in this case is similar to the previous one : The case $q = 0$ correspond to an uncharged scalar field as we already discussed. Here when $q \neq 0$, since $A_H \neq 0$, the system would be subject to electric and magnetic interaction since the black hole as both electric and magnetic charge. For a given value of $A_H$ (and then for a fixed magnetic interaction with the black hole), on most of the allowed parameter space, the magnetic effect is dominant with respect to the electric repulsion. The effect of this magnetic interaction will be to concentrate the scalar field near the horizon. Nevertheless, if one continue to increase $q$, the electric repulsion will finally becomes dominant and lead to a dispersion of the scalar field. The interesting thing here is that this breakdown occur for a value of $q$ independent of 
the parameter of the model, i.e. not only $A_H$ but also $\Omega_H$ and $r_H$. 
This ``universality'' of $q_{\max}$ apears as a surprise and we
did not found any physical or analytic explanation for it.

\begin{figure}[ht!]
\begin{center}
{\includegraphics[width=8cm]{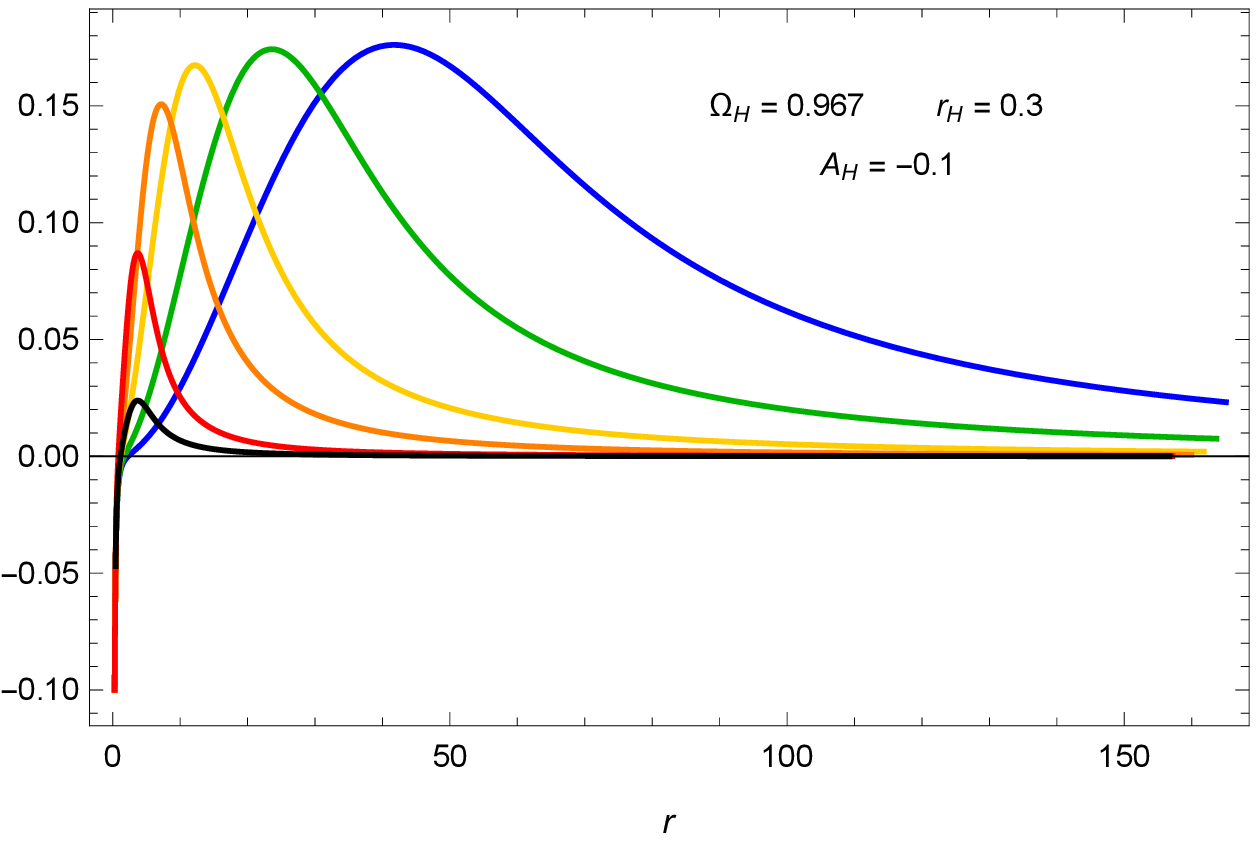}}
{\includegraphics[width=8cm]{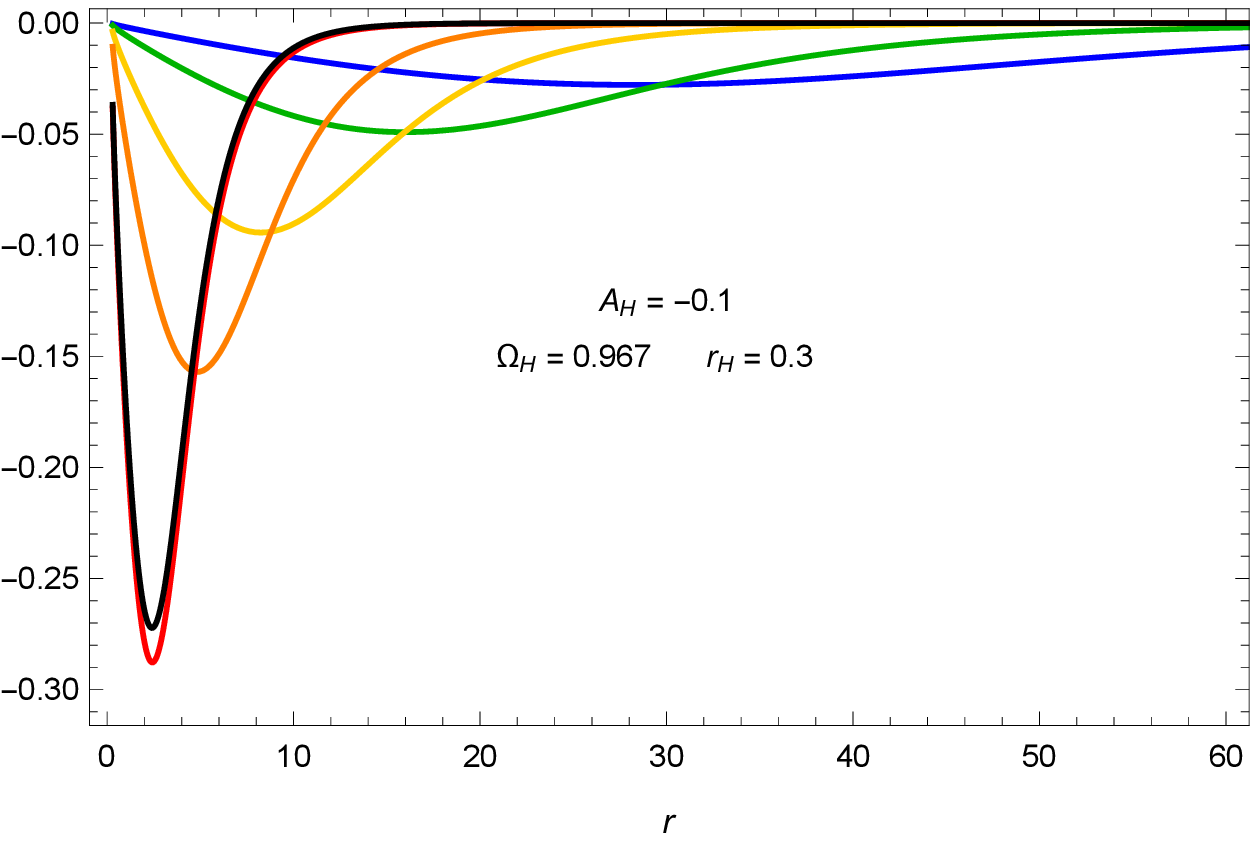}}
\caption{ {\it Left}: 
Profile of the magnetic potential $A$ with $A_H=-0.1$ for $q=0.1, 0.3, 0.5, 0.55, 0.57$ and $0.575\approx q_{\max}$ 
respectively in black, red, orange, yellow, green and blue. These curves are for $\Omega_H=0.967$ and $r_H= 0.3$. 
{\it Right:} Idem for the scalar function $F$.
\label{profile}
}
\end{center}
\end{figure}

Let us finally mention that, when the value $q_{\max}$ is approached,
 the mass and angular momentum of the black hole become very large and possibly tend to infinity
 (see Fig.\ref{data_e}, right side). Due to numerical difficulties this statement can hardly be proven  
 but an inspection of the profiles presented on Fig. \ref{profile}
 show that the fields have tendency to spread over space for $q \to q_{\max}$. We checked that this behaviour persist for different values of $A_H$, $r_H$ and $\Omega_H$ than the ones presented on the figure.

\section{Conclusion}

This paper is devoted to the study of higher dimensional rotating charged black holes in Einstein gravity supplemented by a doublet of complex massive scalar fields (see section \ref{sectmodel} for the details of the model).

This model contain numerous parameters. Within our ansatz, once fixed the units ($8\pi G = 1$ and $\mu = 1$, where $\mu$ is the mass of the scalar doublet), the solutions are determined in terms of 4 parameters : the position of the event horizon $r_H$, the angular velocity of the black hole at the horizon $\Omega_H$, the gauge coupling parameter $q$ and the value of the magnetic potential at the horizon $A_H$.

In sections \ref{sectnonhairy} and \ref{sectunchargedhairy} we reviewed the main results already available in the literature for the non hairy and uncharged hairy sub-sectors of our model.

In section \ref{sectchargedhairy} we finally comes with new results for the most general cases \textit{i.e.} charged rotating hairy solutions. We saw that, any uncharged hairy solution (described by $r_H$ and $\Omega_H$) gives rise to a two-parameter space of solutions (controlled by $q$ and $A_H$). As expected, the domain of existence for the solutions is asymmetric with respect to a change of sign of $A_H$ only when $q \neq 0$. We saw that the domain of existence was more restricted for $A_H > 0$ while for $A_H < 0$ solutions might exists for a significantly higher value of $|A_H|$. In both cases, the bound in the domain is controlled by the condition limiting the possible existence of regular solutions with \textbf{localized} scalar field. That is $\mu - \left\vert\omega - q V_0\right\vert > 0$, where $\omega$ is the harmonic frequency of the scalar doublet and $V_0$ the value of the electric potential at infinity. Basically, this condition fix the maximal value of the gauge coupling parameter, say $q_{\max}$, for which the presence of localized scalar field might be supported by the rotating charged black hole. Our numerical results tend to proof that this value $q_{\max}$ is ``universal'', by this we mean independent of $A_H$, $r_H$ and $\Omega_H$. In our units, we found $q_{\max}\approx0.577$.

Finally, let us mention that the link between the harmonic frequency of the scalar doublet $\omega$ and the angular velocity of the black hole at the horizon $\Omega_H$ was also established and give rise to the condition: $\omega -  \Omega_H  + q A_H \Omega_H = 0$. This condition generalize the synchronization condition known for uncharged solutions.


\clearpage


\begin{thebibliography}{99}


\bibitem{Emparan:2008eg}
  R.~Emparan and H.~S.~Reall,
  Living Rev.\ Rel.\  {\bf 11} (2008) 6.



\bibitem{lovelock}
D. Lovelock, 
J.Math.Phys. 12 (1971) 498-501.






\bibitem{Chrusciel:2012jk}
  P.~T.~Chrusciel, J.~Lopes Costa and M.~Heusler,
  Living Rev.\ Rel.\  {\bf 15} (2012) 7.



\bibitem{Emparan:2001wn}
  R.~Emparan and H.~S.~Reall,
  Phys.\ Rev.\ Lett.\  {\bf 88} (2002) 101101.
  

\bibitem{Ruffini:1971bza}
  R.~Ruffini and J.~A.~Wheeler,
  Phys.\ Today {\bf 24} (1971) no.1,  30.

\bibitem{luckock_moss} 
H. Luckock and I. Moss, 
Phys. Lett. B 176, 341 (1986).

\bibitem{Bekenstein:1996pn}
  J.~D.~Bekenstein,
  ``Black hole hair: 25 - years after,''
  In *Moscow 1996, 2nd International A.D. Sakharov Conference on physics* 216-219
  [gr-qc/9605059].

\bibitem{Herdeiro:2014goa}
  C.~A.~R.~Herdeiro and E.~Radu,
  Phys.\ Rev.\ Lett.\  {\bf 112} (2014) 221101


\bibitem{Herdeiro:2014jaa}
  C.~Herdeiro and E.~Radu,
  Phys.\ Rev.\ D {\bf 89} (2014) no.12,  124018



\bibitem{Herdeiro:2016gxs}
  C.~A.~R.~Herdeiro, E.~Radu and H.~F.~Runarsson,
  Int.\ J.\ Mod.\ Phys.\ D {\bf 25} (2016) no.09,  1641014.


\bibitem{Myers:1986un}
R.~C.~Myers and M.~J.~Perry,
Annals Phys.\  {\bf 172} (1986) 304.

\bibitem{Hartmann:2010pm}
B.~Hartmann, B.~Kleihaus, J.~Kunz and M.~List,
Phys.\ Rev.\ D {\bf 82} (2010) 084022.



\bibitem{Brihaye:2014nba}
  Y.~Brihaye, C.~Herdeiro and E.~Radu,
  Phys.\ Lett.\ B {\bf 739} (2014) 1
\bibitem{Brihaye:2016vkv}
  Y.~Brihaye, C.~Herdeiro and E.~Radu,
  Phys.\ Lett.\ B {\bf 760} (2016) 279
  
\bibitem{Dias:2011at}
  O.~J.~C.~Dias, G.~T.~Horowitz and J.~E.~Santos,
  JHEP {\bf 1107} (2011) 115
\bibitem{Hod:2016kpm}
  S.~Hod,
  Phys.\ Lett.\ B {\bf 755} (2016) 177.

\bibitem{Hod:2017kpt}
  S.~Hod,
  Phys.\ Lett.\ B {\bf 751} (2015) 177



\bibitem{Hod:2013eea}
  S.~Hod,
  Phys.\ Lett.\ B {\bf 713} (2012) 505
\bibitem{Delgado:2016jxq}
  J.~F.~M.~Delgado, C.~A.~R.~Herdeiro, E.~Radu and H.~Runarsson,
  Phys.\ Lett.\ B {\bf 761} (2016) 234






\bibitem{Grandi:2017zgz}
  N.~Grandi and I.~Salazar Landea,
  Phys.\ Rev.\ D {\bf 97} (2018) no.4,  044042.

\bibitem{Brihaye:2018nta}
  Y.~Brihaye and B.~Hartmann,
  ``Critical phenomena of charged Einstein-Gauss-Bonnet black holes with charged scalar hair,''
  arXiv:1804.10536 [gr-qc]; Class. Quant. Grav. (in press).

\bibitem{Brihaye:2018woc}
Y.~Brihaye, T.~Delplace, C.~Herdeiro and E.~Radu,
Phys.\ Lett.\ B {\bf 782} (2018) 124.

\bibitem{COLSYS}
 U. Ascher, J. Christiansen, R.~D. Russell,
 Math. of Comp. {\bf 33} (1979) 659;\\
 U. Ascher, J. Christiansen, R.~D. Russell,
 ACM Trans. {\bf 7} (1981) 209.


\end{thebibliography}
 \end{document}